\documentstyle[aps]{revtex}
\begin{document}
\newcommand{\Kalbermann}{K\"{a}lbermann }

  \newcommand{\asu}{$^1$}
  \newcommand{\iasa}{$^2$}
  \newcommand{\uov}{$^3$}
  \newcommand{\bmit}{$^4$}
  \newcommand{\tjl}{$^5$}
  \newcommand{\wam}{$^6$}
  \newcommand{\uob}{$^7$}
  \newcommand{\uma}{$^8$}
  \newcommand{\uom}{$^9$}
  \newcommand{\fsu}{$^{10}$}
  \newcommand{\odu}{$^{11}$}
  \newcommand{\uoi}{$^{12}$}
  \newcommand{\rpi}{$^{13}$}

\draft

\title{Search for Quadrupole Strength in the Electro-excitation of the
$\Delta^+(1232)$}

\author{
C.~Mertz,$^{1,2} $
C.~Vellidis,\iasa $ $
R.~Alarcon,\asu $ $
D.H.~Barkhuff,\uov$ $
A.M.~Bernstein,\bmit $ $
W.~Bertozzi,\bmit $ $
V.~Burkert,\tjl $ $
J.~Chen,\bmit $ $
J.R.~Comfort,\asu $ $
G.~Dodson,\bmit $ $
S.~Dolfini,\asu$ $
K.~Dow,\bmit$ $
M.~Farkhondeh,\bmit $ $
J.M.~Finn,\wam $ $
S.~Gilad,\bmit $ $
R.W.~Gothe,\uob $ $
X.~Jiang,\uma $ $
K.~Joo,\bmit$ $
N.I.~Kaloskamis,$^{2,4} $
A.~Karabarbounis,\iasa $ $
J.J.~Kelly,\uom $ $
S.~Kowalski,\bmit $ $
C.~Kunz,\uob$ $
R.W.~Lourie,\uov$ $
J.I.~McIntyre,\wam$ $
B.D.~Milbrath,\uov$ $
R.~Miskimen,\uma  $ $
J.H.~Mitchell,\tjl $ $
C.N.~Papanicolas,\iasa $ $
C.F.~Perdrisat,\wam $ $
A.J.~Sarty,\fsu $ $
J.~Shaw,\uma $ $
S.-B.~Soong,\bmit $ $
D.~Tieger,\bmit $ $
C.~Tschal\ae r,\bmit $ $
W.~Turchinetz,\bmit $ $
P.E.~Ulmer,\odu $ $
S.~Van~Verst,\bmit$ $
G.A.~Warren,\bmit$ $
L.B.~Weinstein,\odu $ $
S.~Williamson,\uoi $ $
R.J.~Woo,\wam$ $
A.~Young\asu$ $\\
}

\address{
 {\asu  \it Department of Physics and Astronomy,
                Arizona State University,
                Tempe, Arizona 85287} \\
 {\iasa \it Institute of Accelerating Systems and Applications and
                Department of Physics,
                University of Athens, Athens, Greece} \\
 {\uov \it Institute for Nuclear and Particle Physics
                and Department of Physics,
                University of Virginia,
                Charlottesville, Virginia 22901}  \\
 {\bmit  \it Department of Physics,
                Laboratory for Nuclear Science and Bates Accelerator Center,
                Massachusetts Institute of Technology,
                Cambridge, Massachusetts 02139} \\
 {\tjl \it Thomas Jefferson National Accelerator Facility,
                Newport News, Virginia 23606} \\
 {\wam  \it Physics Department,
                College of William and Mary,
                Williamsburg, Virginia 23187} \\
 {\uob  \it Department of Physics,
                Bonn University,
                Bonn, Germany} \\
 {\uma \it Department of Physics,
                University of Massachusetts,
                Amherst, Massachusetts 01003} \\
 {\uom \it Department of Physics,
                University of Maryland,
                College Park, Maryland 20742} \\
 {\fsu  \it Department of Physics,
                Florida State University,
		Tallahasse, Florida 32306}\\
 {\odu  \it Department of Physics,
                Old Dominion University,
                Norfolk, Virginia 23529} \\
 {\uoi \it Physics Department,
                University of Illinois,
                Urbana-Champaign, Illinois 61801} \\
}

\date{\today}
\maketitle
\begin{abstract}
High-precision H$(e,e^\prime p)\pi^0$ measurements
at $Q^2 = 0.126$ (GeV/c)$^2$ are reported, which allow the determination of
quadrupole amplitudes in the $\gamma^* N\rightarrow \Delta$ transition;
they simultaneously test the reliability of electroproduction models. 
The derived quadrupole-to-dipole amplitude ratios,
$R_{SM}=(-6.5\pm 0.2_{\rm stat+sys}\pm 2.5_{\rm mod})\%$ and
$R_{EM}=(-2.1\pm 0.2_{\rm stat+sys}\pm 2.0_{\rm mod})\%$,
are dominated by model error.
Previous $R_{SM}$ and $R_{EM}$ results should be reconsidered after the model
uncertainties associated with the method of their extraction are taken
into account.

\end{abstract}

\narrowtext
\twocolumn

The conjecture that the nucleon is deformed, raised more than 20 years
ago \cite{glashow}, continues to be  the subject of intense theoretical
\cite{capstick90,isgur82,carlson,buchmann,lu,kamalov,gellas}
and experimental
\cite{blanpied,beck00,batzner,siddle,alder,kalleicher,frolov}
activity.
Because the quadrupole moment of the nucleon vanishes
on account of its spin-1/2 nature, this investigation has naturally turned
to the search for quadrupole strength in the $\gamma^* N\rightarrow
\Delta$(1232) transition.

Spin-parity selection rules in the
$N(J^{\pi}=1/2^+)\rightarrow \Delta(J^{\pi}=3/2^+)$
transition allow magnetic dipole (M1) and  electric (E2)
or Coulomb quadrupole (C2) amplitudes. In the naive (spherical) quark model of
the nucleon, the $\Delta$ excitation is understood as a pure spin-flip
(M1) transition. Experimentally, M1 is indeed found to dominate.
In more refined models, small
E2 and C2 amplitudes are predicted. The physical origin of these
contributions is attributed to different mechanisms in the various models.
However
they invariably have important implications for our understanding
of the structure of the nucleon and of QCD at low energies
\cite{capstick90,isgur82,carlson,buchmann}.

In pion production, the multipole amplitudes are denoted by
$M^{I}_{l\pm }$, $E^{I}_{l\pm }$, and $S^{I}_{l\pm }$,
indicating their character (magnetic, electric, or scalar),
their isospin (I), and their total angular momentum
$(J= l\pm 1/2)$.
Thus, the resonant photon multipoles M1, E2, and C2
correspond to
$M^{3/2}_{1+}$, $E^{3/2}_{1+}$, and $S^{3/2}_{1+}$, respectively.
The Electric- and Scalar-to-Magnetic-Amplitude-Ratio are defined as
$R_{EM} = \Re e(E^{3/2}_{1+}/M^{3/2}_{1+})$
and $R_{SM} = \Re e(S^{3/2}_{1+}/M^{3/2}_{1+})$
respectively.  Most models of the nucleon have definite predictions
for these ratios. They are  invariably very small at low momentum transfers,
the domain of the reported measurements.  The predictions for $R_{EM}$
at $Q^2 = 0$ range from $-$0.1\% up to $-$5\%
\cite{capstick90,isgur82,carlson,buchmann}.

While $R_{EM}$ measurements at $Q^2 = 0$ are pursued with the use
of real photons, its $Q^2$ evolution and the $R_{SM}$ ratio can
be investigated only through electro-excitation.
A number of calculations explore the dependence of $R_{EM}$ and $R_{SM}$
on $Q^2$ \cite{capstick90,lu,kamalov,gellas}.
The experimental determination of $R_{EM}$ and $R_{SM}$ is severely
complicated by the presence of non-resonant processes that are coherent
with the resonant excitation of the $\Delta(1232)$ \cite{pa89}.
These processes (such as Born contributions and tails of higher resonances),
termed ``background contributions,''
need to be constrained with model calculations and measurements tailored to
this end. Also, it is imperative that electro-production models, used in model
extraction of $R_{EM}$ and $R_{SM}$, are adequately tested in their ability to
accurately handle small amplitudes, both resonant and background.

Precision measurements with polarized tagged
photons have resulted in an
$R_{EM}$ at resonance of
$(-3.0 \pm 0.3)\%$ \cite{blanpied} and $(-2.5 \pm 0.3) \%$ \cite{beck00}.
Model calculations are in reasonably good agreement with
experiment \cite{drechsel98,sato_and_lee,dm}.
The situation is quite different for electron scattering investigations.
Experiments conducted
in the late 60's and early 70's for  $Q^2$ up to 1 (GeV/c)$^2$
have yielded $R_{EM}$ values consistent with zero and $R_{SM}$
of around $-$7\% with large statistical
and systematic errors \cite{batzner,siddle,alder}.
A dispersion relation analysis~\cite{crawford} reported
exceptionally large values of $R_{SM}$ around $-13\%$
in the range of $Q^2 = 0.1 - 0.25$ (GeV/c)$^2$,
suggestive of a narrow structure
peaking near $Q^2= 0.1$ (GeV/c)$^2$.
These values are consistent with the value
$\Re e(S_{1+}/M_{1+}) = (-12.7 \pm 1.5)\%$ of the ratio of isospin-mixed
multipoles which was reported in a recent
H$(e,e^\prime \pi^0)p$ experiment at $Q^2 = 0.127$ (GeV/c)$^2$
\cite{kalleicher}.
The measurements reported here, performed at the same $Q^2$,
allow a direct comparison with the afore-mentioned data.

The coincident H$(e,e^{\prime}p)\pi^{0}$ cross section in the
One-Photon-Exchange-Approximation can be written as~\cite{drechsel}:

\begin{eqnarray}
\frac{d\sigma}{d\omega d\Omega_e d\Omega^{cm}_{pq}} &=&
\Gamma_v~\frac{p_{cm}}{k_{cm}}~\sigma~,              \\
\nonumber
\sigma =
R_{T} + \varepsilon_{L} R_{L} &-&
\rho_{LT}R_{LT}\cos\phi
+ \varepsilon R_{TT}\cos 2\phi~,
\label{eq:xsection}
\end{eqnarray}

\noindent
where $\Gamma_v$ is the virtual photon flux;
$p_{cm}$ and $k_{cm}$ are the pion momentum and the
photon equivalent energy in the hadronic CM frame, respectively;
$\varepsilon$,
$\varepsilon_{L}$,
and $\rho_{LT}$
are electron kinematic factors;
and $\phi$ is the nucleon azimuthal angle about the momentum transfer
$\vec{q}$ measured from the nucleon direction closest to the beam exit line.
$R_{L}$, $R_{T}$, $R_{LT}$, and $R_{TT}$ are the
longitudinal, transverse, longitudinal-transverse, and transverse-transverse
interference response functions, respectively~\cite{drechsel}.

To study the $\gamma^* N\rightarrow\Delta$ transition with high precision,
an extensive program has been developed at the MIT-Bates
Linear Accelerator.
We report here %
results from the first phase of the program.
We have reported the recoil proton polarization $P_n$ result
from the same experiment \cite{warren}.

The experiment \cite{warren,mertz} was conducted
at energies of 719  and 799 MeV and
a liquid H$_{2}$ target was used;
the scattered electrons were detected in the ``MEPS'' spectrometer
and the coincident protons in ``OHIPS''. %
The focal plane instrumentation of each spectrometer
consisted of one crossed vertical drift chamber for track reconstruction
and scintillators for triggering.
Detailed optics studies were done for each spectrometer, and the detection
efficiencies were measured as functions of all independent reaction
coordinates. 
The phase-space normalization of the cross section and various corrections
applied to the data,  including radiative corrections,
were implemented with the aid of a Monte Carlo simulation model.
The coincident cross section was measured at $\phi = 0$ and $\pi$
for a broad range of hadronic mass W
around the resonance and a range of proton polar angle $\theta$ about
$\vec{q}$ in the hadronic CM frame near $\theta=0$.

\begin{figure}
\vspace{3.8 in}
\includegraphics{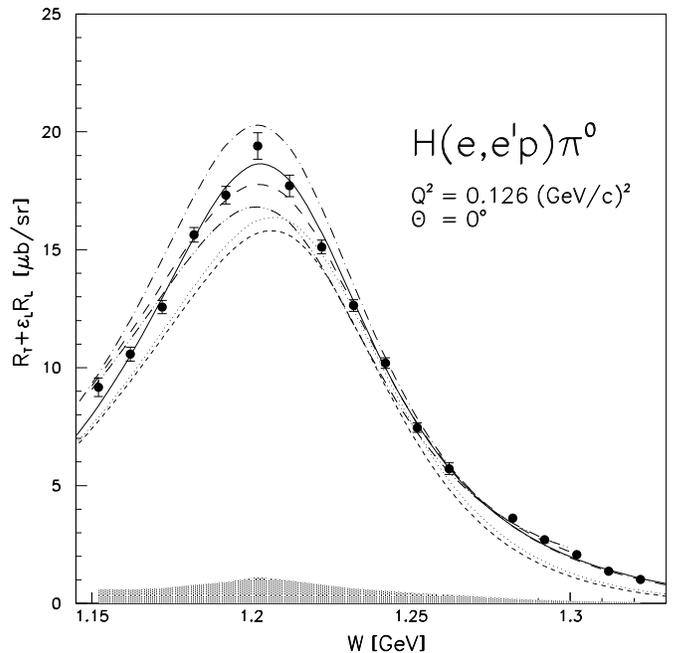}
\caption{The CM cross section in parallel kinematics.
The solid curve is the fit of MAID [17,24] and the dot-dashed curve is the
corresponding result for the resonant E2$=$C2$=$0.
The long-dashed curve is the fit of RPI [19,26] and the dot-dot-dashed curve
is the corresponding result for the resonant C2$=$0.
The short-dashed and dotted curves are the ``deformed'' and ``non-deformed''
prediction of SL [18,25], respectively.
The shaded band depicts the value of the systematic error.}
\label{fig:sigma0_w}
\end{figure}

\begin{figure}
\vspace{3.9 in}
\includegraphics{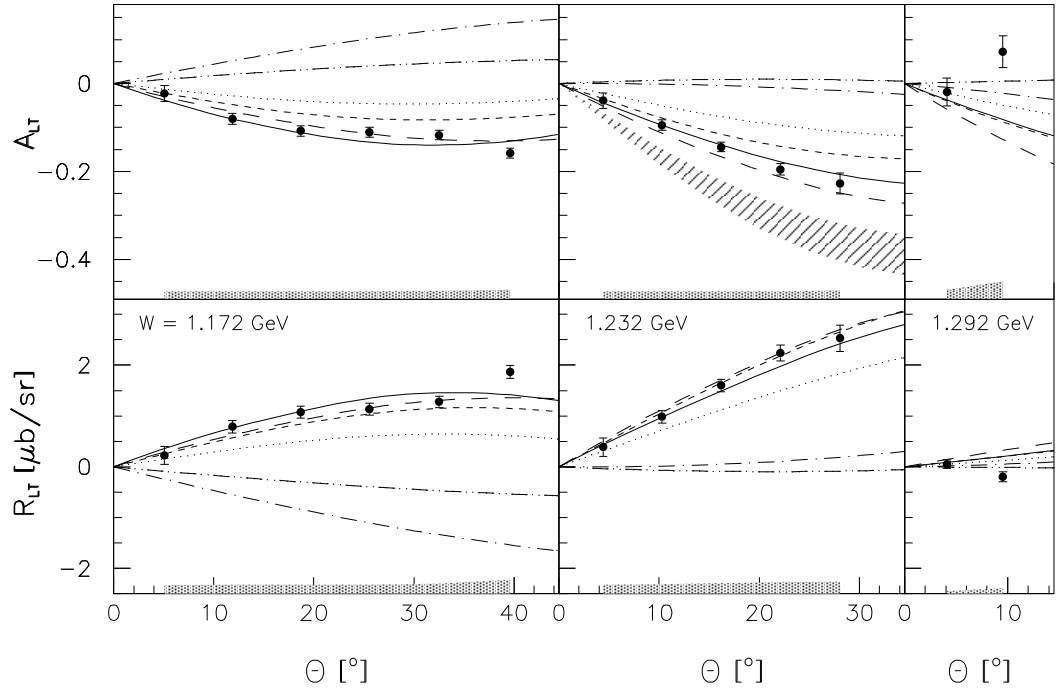}
\caption{The longitudinal-transverse asymmetry and response,
as a function of the proton polar angle relative to $\vec{q}_{cm}$.
The
curves are explained in Fig.\ \ref{fig:sigma0_w}.
The hatched band is the projection of the Bonn result [14],
and the shaded bands depict the values of  the  systematic error.}
\label{fig:asy}
\end{figure}

Fig.\ \ref{fig:sigma0_w} shows the coincident cross section as a function of
the hadronic mass W for proton detection
at $\theta=0$,
where $R_{LT}$ and $R_{TT}$ vanish
and $R_{T}$ has the maximum sensitivity to $\Re e(E_{1+}^{*}M_{1+})$.
The data exhibit a distinct resonant shape, arising mostly from
$\vert M_{1+}\vert^{2}$.
Fig.\ \ref{fig:asy} shows the response function $R_{LT}$ and the cross section
asymmetry $A_{LT}$ which are sensitive to $\Re e(S_{1+}^{*}M_{1+})$,

\begin{equation}
A_{LT} = \frac{\sigma_{\phi = 0} - \sigma_{\phi = \pi}}
              {\sigma_{\phi = 0} + \sigma_{\phi = \pi}}
       = \frac{-\rho_{LT}R_{LT}}{R_{T} + \varepsilon_{L} R_{L} +
                                         \varepsilon R_{TT}}~.
\label{eq:alt}
\end{equation}

The measured cross section (Fig.\ \ref{fig:sigma0_w}), asymmetry and $R_{LT}$
response function (Fig.\ \ref{fig:asy}),
are compared with the curves that result by adjusting the relevant parameters
in the models of of Drechsel {\it et al.}
\cite{drechsel98,maid} (MAID),
of Sato and Lee (SL)~\cite{sato_and_lee,private_lee},
and of Davidson and Mukhopadhyay (RPI)~\cite{dm,david}.
All three models start from the same Lagrangian for the non-resonant terms,
including explicit nucleon and light meson ($\pi$, $\rho$, $\omega$) degrees
of freedom coupled to the electromagnetic field. Their principal differences
lie in the definition of the $\Delta$ resonance and in the method of
unitarization. The solid curve is the fit of the MAID-2000 code, where the five
parameters controlling the electromagnetic couplings of the $\Delta$(1232)
and of the P$_{11}$(1440) (Roper) resonances were fitted to our data.
The long-dashed curve results by adjusting all seven free parameters
of the RPI model, in a fashion similar to that reported in \cite{frolov},
while the short-dashed curve results by judiciously adjusting
(without $\chi^2$ minimization) the parameters of the SL model to the data.
All calculations properly obtain the position of the cross section
maximum. They differ in their detailed shape and in magnitude. The
adjusted MAID-2000 and RPI models provide an excellent description
of the data, with the possible exception of the high W points.

The $A_{LT}$ and $R_{LT}$ results (Fig.\ \ref{fig:asy}) amply demonstrate the
sensitivity of our data to the presence of resonant quadrupole amplitudes.
All three models fail dramatically
if the resonant quadrupole amplitudes are
set to zero.
However, when the quadrupole strength is
adjusted,
good agreement is achieved.

The sensitivity of our data to the quadrupole amplitudes allows for the
determination of $R_{EM}$ and $R_{SM}$ either through a variant of the
M1-dominance truncated multipole expansion (``TME'') fit (as in
\cite{batzner,siddle,alder,kalleicher}) or through model extraction, as in
\cite{frolov}. The derived values are shown in Table \ref{tab:multipoles}.
In TME fit (a), as in \cite{kalleicher},
it is assumed that only
the multipoles $M_{1+}$ and $S_{1+}$ contribute and that $R_L$
is insignificant.
Then we obtain
$\Re e(S_{1+}/M_{1+})=(-7.6\pm 0.3_{\rm stat}\pm 0.7_{\rm sys})\%$.
The hatched band in Fig.\ \ref{fig:asy}
shows the projected asymmetry ($1\sigma$ confidence) for our angular range
if the Bonn
$\Re e(S_{1+}/M_{1+})=(-12.7\pm 1.5_{\rm stat})\%$~\cite{kalleicher} is
adopted. Our data points lie several standard deviations away.
Noting that $A_{LT}$ was measured in~\cite{kalleicher} near
$\theta=180^{\circ}$, the discrepancy may indicate that terms having
a different dependence on $\theta$ than those included in
TME fit (a) contribute significantly. If all three $1^{+}$
multipoles are adjusted, setting
$\vert S_{1+}\vert^{2}$=$\vert E_{1+}\vert^{2}$=$\Re e(S_{1+}^{*}E_{1+}$)=0,
the derived value of $\Re e(S_{1+}/M_{1+})$, labeled with ``TME (b)''
in Table \ref{tab:multipoles}, is noticeably larger, although not incompatible
with, the value extracted through
TME fit (a). This is a manifestation of the significant truncation error
that characterizes the TME approach.

$\Re e(S_{1+}/M_{1+})$ and $\Re e(E_{1+}/M_{1+})$
values are also obtained from the fits of the MAID-2000~\cite{maid}
and RPI~\cite{david} models
and from the adjustment of the SL model~\cite{private_lee}.
While all three models achieve
a reasonable agreement with the unpolarized data (Figs. \ref{fig:sigma0_w} and
\ref{fig:asy}), the resulting values of $P_n$ disagree with each
other (Table \ref{tab:multipoles}) %
and with the experimental value
$P_n = -0.397\pm 0.055_{\rm stat}\pm 0.009_{\rm sys}$~\cite{warren}.
The MAID-2000 value could be considered as providing a fair agreement, lying
within two standard deviations from the experimental value.

Given the overall success of the MAID-2000 model
fit in accounting for our data,
we adopt the values $R_{SM}=(-6.5\pm 0.2_{\rm stat+sys})\%$,
$R_{EM}=(-2.1\pm 0.2_{\rm stat+sys})\%$, and
$\vert M_{1+}^{3/2}\vert=(39.8\pm 0.3_{\rm stat+sys})
\times 10^{-3}/{m}_{\pi^{+}}$.
The statistically incompatible values provided by the other two,
equally sophisticated, model analyses indicate that the results are
characterized by substantial model uncertainty. 
The quantification of this uncertainty for each one of the available models
is urgently needed. It could remove the apparent contradictions among the
available models.
We assume that the scatter of the extracted
values provides an estimate of the model uncertainty.
We therefore attribute, conservatively, to $R_{SM}$ 
and $R_{EM}$ 
model uncertainties of $\pm 2.5\%$ and $\pm 2.0\%$ respectively,
and to $\vert M_{1+}^{3/2}\vert$ a model uncertainty of
$\pm 2.0\times 10^{-3}/{m}_{\pi^{+}}$.
Previously published $R_{SM}$ and $R_{EM}$ results
\cite{batzner,siddle,alder,kalleicher,frolov} have not taken into account
this uncertainty. They are subject to comparable model error. This added
uncertainty may remove all known inconsistencies amongst them, when properly
estimated.

The data presented here exhibit unprecedented
sensitivity 
to the presence of resonant quadrupole amplitudes. 
Their analysis leads us to the following conclusions:
i) Extractions of quadrupole strengths based on 
TME fits are characterized by substantial truncation error
and lead to inconsistent results;  ii) Claims of large 
$R_{SM}$ at low $Q^2$
derived from earlier~\cite{crawford} and recent~\cite{kalleicher}
measurements cannot be supported; 
iii) Even when conservative estimates of systematic and model uncertainties
are taken into consideration, an unambiguously negative value for $R_{SM}$
is obtained. This value supports the claims for an oblate deformed $\Delta$;
and iv) %
The available pertinent electroproduction models
are on the verge of succesfully describing the high precision data that are
now emerging.
It is important that the model errors due to input parameters and
model assumptions be quantified. It is essential that 
measurements be performed that are sensitive to background amplitudes,
along with those that are primarily sensitive to quadrupole amplitudes.

We are indebted to Drs. S.S.~Kamalov, D.~Drechsel, L.~Tiator, R.M.~Davidson,
N.C.~Mukhopadhyay, T.-S.H.~Lee, T.~Sato, and J.M.~Laget for
providing us with detailed calculations and valuable comments concerning
their models and the
issue of ``nucleon deformation.''

\begin{table}\centering
\begin{tabular}{|c|c|c|c|c|}
 Model & $\vert M_{1+}\vert$ & $\Re e(E_{1+}$/M$_{1+}$)
 & $\Re e(S_{1+}/M_{1+}$) & $P_{n}$ \\
 & $(10^{-3}/{m}_{\pi^{+}})$ & (\%) & (\%) & \\\hline
 TME (a) & $25.1\pm 0.7$ & 0 & $-7.6\pm 0.8$ & --- \\
 TME (b) & $24.5\pm 1.1$ & $+0.9\pm 1.4$ & $-8.5\pm 1.2$ & --- \\
 RPI & $25.4\pm 0.3$ & $+0.8\pm 0.8$ & $-9.1\pm 0.8$ & $-0.12$ \\
 MAID & $26.6\pm 0.2$ & $-2.2\pm 0.2$ & $-6.7\pm 0.2$ & $-0.51$ \\
 SL & 27.7 & $-3.3$ & $-4.3$ & $-0.26$ \\\hline
\end{tabular}
\vspace{12pt}
\caption{Multipoles extracted from the present
data at $Q^{2}$$=$0.126 (GeV/c)$^{2}$ and W$=$1232 MeV. 
Statistical and systematic errors are added quadratically.
The measured $P_{n}$=$-0.40\pm 0.06$ at W$=$1231 MeV [22].}
\label{tab:multipoles}
\end{table}

\newpage
{\tiny .}

\newpage
{\tiny .}

\begin{references}
%


\bibitem{glashow}
S.~L. Glashow, Physica {\bf 96A},  27  (1979).

%
%

\bibitem{capstick90}
S. Capstick and G. Karl, Phys. Rev. {\bf D41},  2767  (1990);
%
%
%
S. Capstick and B.~D. Keister, Phys. Rev. {\bf D51},  3598  (1995).

%
%
%

\bibitem{isgur82}
G.~K. N.~Isgur {\it et~al.},
%
Phys. Rev. {\bf D25},  2394  (1982).

%
%

%
%

%
%
%

%
%
%

%
%

%
%

%
%

\bibitem{carlson}
C.~E. Carlson, Phys. Rev. {\bf D34},  2704  (1986).

%
%

%
%

\bibitem{buchmann}
A.~J. Buchmann {\it et~al.},
%
Phys. Rev. {\bf C55},  448  (1997).

%
%

%
%
%

%
%

%
%

\bibitem{lu}
D. H. Lu {\it et~al.}, Phys. Lett. {\bf C55},  3108  (1997).

\bibitem{kamalov}
S.~S. Kamalov and S.~N. Yang, Phys. Rev. Lett {\bf 83}, 4494 (1999).

\bibitem{gellas}
G.~C. Gellas {\it et~al.}, Phys. Rev. {\bf D60}, 54022 (1999).

\bibitem{blanpied}
G. Blanpied {\it et~al.}, Phys. Rev. Lett. {\bf 79},  4337  (1997).

%
%

\bibitem{beck00}
R. Beck {\it et~al.}, Phys. Rev. {\bf C61},  35204  (2000).

\bibitem{batzner}
K. Batzner {\it et~al.}, Nucl. Phys. {\bf B76},  1  (1974).

\bibitem{siddle}
R. Siddle {\it et~al.}, Nucl. Phys. {\bf B35},  93  (1971).

\bibitem{alder}
J.~C. Alder {\it et~al.}, Nucl. Phys. {\bf B46},  573  (1972).

\bibitem{kalleicher}
F. Kalleicher {\it et~al.}, Z. Phys. {\bf A359},  201  (1997).

\bibitem{frolov}
V.~V. Frolov {\it et~al.}, Phys. Rev. Lett. {\bf 82}, 45 (1999).

\bibitem{pa89}
C. N. Papanicolas, in {\it Topical Workshop on Excited Baryons},
%
World Scientific, Singapore, 1989;
A.M. Bernstein {\it et~al.}, Phys. Rev. {\bf C47},  1274  (1993).

%
%

%
%

\bibitem{drechsel98}
D. Drechsel {\it et~al.}, Nucl. Phys. {\bf A645}, 145 (1999).

\bibitem{sato_and_lee}
T. Sato and T.-S.~H. Lee, Phys. Rev. {\bf C54},  2660  (1996).

\bibitem{dm}
%
%
%
R.~M. Davidson {\it et~al.}, Phys. Rev., {\bf D43}, 71 (1991);
%
R.~M. Davidson and N.~C. Mukhopadhyay, Phys. Lett. {\bf B353}, 131 (1995).

\bibitem{crawford}
R.~L. Crawford, Nucl. Phys. {\bf B28},  573  (1971).

%
%

\bibitem{drechsel}
D. Drechsel and L. Tiator,
%
J. Phys. G: Nucl. Part. Phys. {\bf 18},  449  (1992).
%
%
%
%
%

%
%
%
%
%
%
%
%
%
%
%
%

\bibitem{warren}
G.~A. Warren {\it et~al.}, Phys. Rev. {\bf C58}, 3722 (1998).

\bibitem{mertz}
C. Mertz, PhD thesis, Arizona State University, unpublished (1998);
%
%
C. Vellidis, PhD thesis, University of Athens, in preparation.

%
%
%

%
%
%

%
%
%
%
%

%
%

%
%

\bibitem{maid}
S.~S. Kamalov, private communication, 2000.

\bibitem{private_lee}
T. Sato and T.-S.~H. Lee, private communication, 2000, and to be published.

\bibitem{david}
R.~M. Davidson, private communication, 2000.


\end{references}
\end{document}